\documentclass[conference]{IEEEtran}
\IEEEoverridecommandlockouts

\usepackage{amsmath,url}

\usepackage{cite}
\usepackage{amsmath,amssymb,amsfonts, bm}
\newcommand{\R}{\mathbb{R}}
\usepackage{enumerate}
\usepackage[linesnumbered,ruled,vlined]{algorithm2e}
\let\oldnl\nl
\newcommand{\nonl}{\renewcommand{\nl}{\let\nl\oldnl}}

\usepackage[]{graphicx}
\graphicspath{{Figures/}}
\usepackage{support-caption}
\usepackage{subcaption}
\usepackage{lipsum}
\usepackage{balance}
\usepackage{textcomp}
\usepackage{xcolor}
\usepackage{multirow}
\usepackage{color}
\usepackage{xcolor}
\usepackage{listings}
\usepackage{courier}
\usepackage{float}
\usepackage{verbatim}
\usepackage{stfloats}
\usepackage{pifont}
\newcommand{\cmark}{\textcolor{red}{\ding{51}}}%
\newcommand{\xmark}{\textcolor{green}{\ding{55}}}%
\usepackage{array}
\usepackage{comment}

\usepackage{fancyhdr}
\definecolor{CustomColor}{RGB}{255, 255,255}

\lstdefinelanguage{turtle}
{
    columns=fullflexible,
    keywordstyle=\color{red},
    morekeywords={@prefix,@base,@forSome,@forAll,@keywords},
    morecomment=[l]{\#},
    tabsize=4, 
    alsoletter={-?}, 
    morecomment=[s][\color{blue}]{<}{>},
    basicstyle=\ttfamily\color{black}, 
    morestring=[b][\color{black}]\",    
    backgroundcolor=\color{CustomColor}
}
\lstdefinestyle{turtle}{%
    morekeywords={a, @prefix},
    morecomment=[s][\rmfamily]{<}{>},
    morecomment=[s][\itshape]{"}{"},
}

\def\BibTeX{{\rm B\kern-.05em{\sc i\kern-.025em b}\kern-.08em
    T\kern-.1667em\lower.7ex\hbox{E}\kern-.125emX}}

\usepackage{nomencl}
\makenomenclature

\usepackage{amsthm}

\newtheorem*{remarkA}{Null hypothesis $H^a_0$}
\newtheorem*{remarkB}{Null hypothesis $H^b_0$}
\newtheorem*{remarkC}{Null hypothesis $H^c_0$}

\makeatletter
\newcommand{\linebreakand}{%
  \end{@IEEEauthorhalign}
  \hfill\mbox{}\par
  \mbox{}\hfill\begin{@IEEEauthorhalign}
}
\makeatother

\begin{document}

\title{Impact of Load Demand Dataset Characteristics \\on Clustering Validation Indices\\\thanks{The  ADAPT  Centre  for  Digital  Content  Technology  is  funded  under  the  SFI Research Centres Programme (Grant 13/RC/2106) and is co-funded under the European Regional Development Fund.}
\thanks{Send correspondence to S.\ Dev, email: soumyabrata.dev@ucd.ie}
\thanks{$^{*}$~Authors contributed equally}
}

\author{\IEEEauthorblockN{Mayank~Jain$^{*}$}
\IEEEauthorblockA{\textit{ADAPT SFI Research Centre} \\
\textit{University College Dublin (UCD)}\\
Dublin, Ireland\\
mayank.jain@adaptcentre.ie}
\and
\IEEEauthorblockN{Mukta~Jain$^{*}$}
\IEEEauthorblockA{\textit{Delhi School of Economics} \\
\textit{University of Delhi (DU)}\\
Delhi, India\\
muktajain@econdse.org}
\linebreakand
\IEEEauthorblockN{Tarek~AlSkaif}
\IEEEauthorblockA{\textit{Information Technology Group} \\
\textit{Wageningen University and Research}\\
Wageningen, The Netherlands\\
tarek.alskaif@wur.nl}
\and 
\IEEEauthorblockN{Soumyabrata~Dev}
\IEEEauthorblockA{\textit{ADAPT SFI Research Centre} \\
\textit{University College Dublin (UCD)}\\
Dublin, Ireland\\
soumyabrata.dev@ucd.ie}
}

\IEEEoverridecommandlockouts

\maketitle
\thispagestyle{fancy}%


\begin{abstract}
With the inclusion of smart meters, electricity load consumption data can be fetched for individual consumer buildings at high temporal resolutions. Availability of such data has made it possible to study daily load demand profiles of the households. Clustering households based on their demand profiles is one of the primary, yet a key component of such analysis. While many clustering algorithms/frameworks can be deployed to perform clustering, they usually generate very different clusters. In order to identify the best clustering results, various cluster validation indices (CVIs) have been proposed in the literature. However, it has been noticed that different CVIs often recommend different algorithms. This leads to the problem of identifying the most suitable CVI for a given dataset. Responding to the problem, this paper shows how the recommendations of validation indices are influenced by different data characteristics that might be present in a typical residential load demand dataset. Furthermore, the paper identifies the features of data that prefer/prohibit the use of a particular cluster validation index.
\end{abstract}

\begin{IEEEkeywords}
clustering, validation indices, residential load demand
\end{IEEEkeywords}

\vspace{-0.3cm}
\section{Introduction}
\label{sec:intro}
An informed production decision by the supplier is crucial for maximizing efficiency in the electricity distribution process. Such a decision requires information about the electricity consumption patterns (or load demand). In industrial, commercial, and transportation sectors, such patterns mostly follow a definite trend and hence are easier to identify. However, it is not the case with the residential sector which has been noted to consume more than $21\%$ of the net electricity consumption~\cite{eia2018}. 

Load demand profiles can now be obtained for individual households at high temporal resolution with the help of smart meters. To make the most from such high-resolution data with varying consumption patterns of individual households, clustering is usually done while pre-processing. The general idea is to analyze households with different consumption patterns separately. An effective clustering procedure thus reduces the net variance of the data and streamlines further analysis. Such clustering helps electricity suppliers in delivering more responsive demand and planning appropriate tariff schemes~\cite{satremeloy2020cluster}.

Multiple clustering strategies have been explored in the literature to study the residential load demand profiles. Jain~\textit{et~al.}~\cite{jain2020clustering} defines such strategies in a generalized two-step clustering framework, where the first step is to reduce dimensionality and the second step comprises of an unsupervised clustering algorithm. Dimensionality reduction is considered useful in the prevention of over-fitting while clustering~\cite{andrews2018addressing} and in complexity reduction of the problem~\cite{jain2021validation}. Yet it remains an optional step that has been omitted in various studies~\cite{al2016, mcloughlin2015}. Nonetheless, multiple combinations of clustering algorithms and dimensionality reduction techniques can be used to perform clustering on electricity load profiles. It has been further noted that different combinations of such algorithms often produce very different clustering results~\cite{williams2013}.

Different cluster validation strategies have been proposed to identify the best algorithm and hence the finest clusters. But, all the strategies themselves have their shortcomings and biases. Therefore, different strategies recommend different algorithms for the same dataset~\cite{jain2021validation, williams2013, liu2010understanding}. It complicates the judgment to find the best clustering. This paper tries to analyze the response of different cluster validation strategies upon change in the characteristics of residential load demand dataset and recommends the preferred cluster validation strategy for different data characteristics.  

\subsection{Relevant Literature}
There is a vast literature in the domain of clustering algorithms. One of the most popular one is the k-means clustering algorithm which uses iterative method to generate clusters~\cite{al2016,mcloughlin2015,yildiz2018,satremeloy2020cluster}. Hierarchical or agglomerative clustering is another highly exploited candidate that starts with considering each point as a separate cluster and proceeds by merging similar ones~\cite{williams2013,granell2014,satremeloy2020cluster}. Many other studies have used fuzzy c-means algorithm (FCM)~\cite{jain2021validation,viegas2015,zhou2017household} which generates clusters with fuzzy boundaries and hence works best in cases where overlapping clusters are observed. Some of the other algorithms that have been considered to cluster residential load demand profiles are self-organizing maps~\cite{mcloughlin2015,yildiz2018} and the Dirichlet Process Mixture Model~\cite{granell2014}. Some studies have also highlighted the importance of dimensionality reduction before clustering the load demand profiles. Most commonly used algorithm is this category is the principal component analysis (PCA). It tries to retain maximum information (explained variance) in minimum number of dimensions~\cite{yildiz2018,jain2021validation}. A few other studies have also considered feature agglomeration as an alternative to PCA~\cite{jain2020clustering}.

In the absence of ground truth labels of underlying clusters, clustering validation indices (CVI) have been defined in the literature to validate the clustering results. The literature presents many CVIs like R-squared measure, SD Index, I index~\cite{maulik2002performance} \textit{etc.} The most widely used strategies in the domain of residential load demand profiles are Silhouette score (SH)~\cite{al2016,yildiz2018,satremeloy2020cluster}, Davies-Bouldin score (DB)~\cite{mcloughlin2015,yildiz2018,satremeloy2020cluster}, Calinski-Harabasz score (CH)~\cite{satremeloy2020cluster}, Xie-Beni index (XB)~\cite{zhou2017household,jain2021validation} and Dunn's index (DI)~\cite{al2016,granell2014}.

It often happens that different CVIs prefer different algorithms~\cite{jain2021validation,williams2013}. To resolve this issue of disparity, researchers have resorted to different statistical strategies. Some took a majority vote to identify optimal clustering~\cite{viegas2015}, while others chose CVI exhibiting minimum variability across different clustering results~\cite{satremeloy2020cluster}. Although the aforementioned strategies might work in some particular cases, a systematic analysis of the impact of data characteristics on the performance of different CVIs is required. To this end, Liu~\textit{et~al.}~\cite{liu2010understanding} studied the validation properties of different CVIs using different experiments on an entirely synthetic data set. However, such dataset is far from the data available in real world and hence its results are not directly comprehensible in practical situations.

\subsection{Contributions and Outline of the Paper}
While most CVIs use `separation' and `compactness' of the resulting clusters in their definition, they adopt significantly different strategies to achieve it~\cite{liu2010understanding,jain2021validation}. Hence, a valid comparative study for the effectiveness of different CVIs can only be conducted by evaluating their response to the change of data characteristics. In such regard, this paper\footnote{With the spirit of reproducible research, the code to reproduce the simulations in this article is shared at \url{https://github.com/jain15mayank/data-characteristics-impact-on-CVIs}} makes the following contributions:
\begin{itemize}
    \item Evaluating the effect of following data (or underlying cluster) characteristics on different CVI scores
    \begin{itemize}
        \item Outliers
        \item Density
        \item Diameter
    \end{itemize}
    \item Recommending the preference/proscription of a CVI if such characteristics are present in the data
\end{itemize}

The rest of the paper is organized as follows. 
Section II explains the data set and its pre-processing. Section III develops the experimental setup that computes baseline clustering and validation indices. Section IV formulates the hypothesis testing procedures while the results are discussed in section V. Finally, section VI concludes the paper.

\section{Dataset and Pre-processing}
\label{sec:pre-processing}
This research is built on the data of real electricity demand profiles. The data is taken from Pecan Street Dataport~\cite{street2015dataport} which is available in public domain for research purposes. The objective of this dataset providers is to boost the research in the domain of electricity load demand profile analysis. The paper considers a sample of $50$ households from New York and Austin ($25$ from each city). The electricity consumption data is provided with a temporal resolution of $15$ minutes in the dataset. While the Austin data is available for $1$ year, New York data is available for roughly $6$ months only. Since the study focuses on clustering of median daily load profiles only and is not considering seasonality and other periodic factors, the data from both cities was merged together to increase the number of data points. This was essential to effectively study the impact of load demand dataset characteristics on the performance of clustering validation indices.

The greater variance and outlier existence in the data of each household necessitates median to be the measure of central tendency to define daily demand profiles of the household~\cite{jain2020clustering}. Afterwards, $\ell_2$ normalization technique is used to identify the trends of electricity demanded by different households.

\section{Baseline Setup}\label{sec:experimentalSetup}
The residential load demand profile data is very peculiar. The profiles are not random but have certain specific trends and distribution. Therefore, the results of synthetic data are inapplicable to these profiles. Referencing of the demand data is necessary for carrying out the analysis. Additionally, the ground truth clusters must be known to study the effect of change in data characteristics. Hence, the baseline setup to take the reference ground truth is carried out in two major stages. The first stage is to form clusters and the second is to compute the CVIs.

\subsection{Clustering Framework}\label{sec:clusteringFramework}

Although studying different clustering methods to perform optimal clustering is an interesting problem, such analysis is beyond the scope of this paper. Hence, the paper performs PCA to reduce the dimensionality and then applies FCM algorithm to perform final clustering of the households. This particular combination was chosen in accordance with the results obtained in a more recent study on clustering residential load demand profiles~\cite{jain2021validation}.

\begin{figure}[!ht]
    \centering
    \includegraphics[width=0.95\columnwidth]{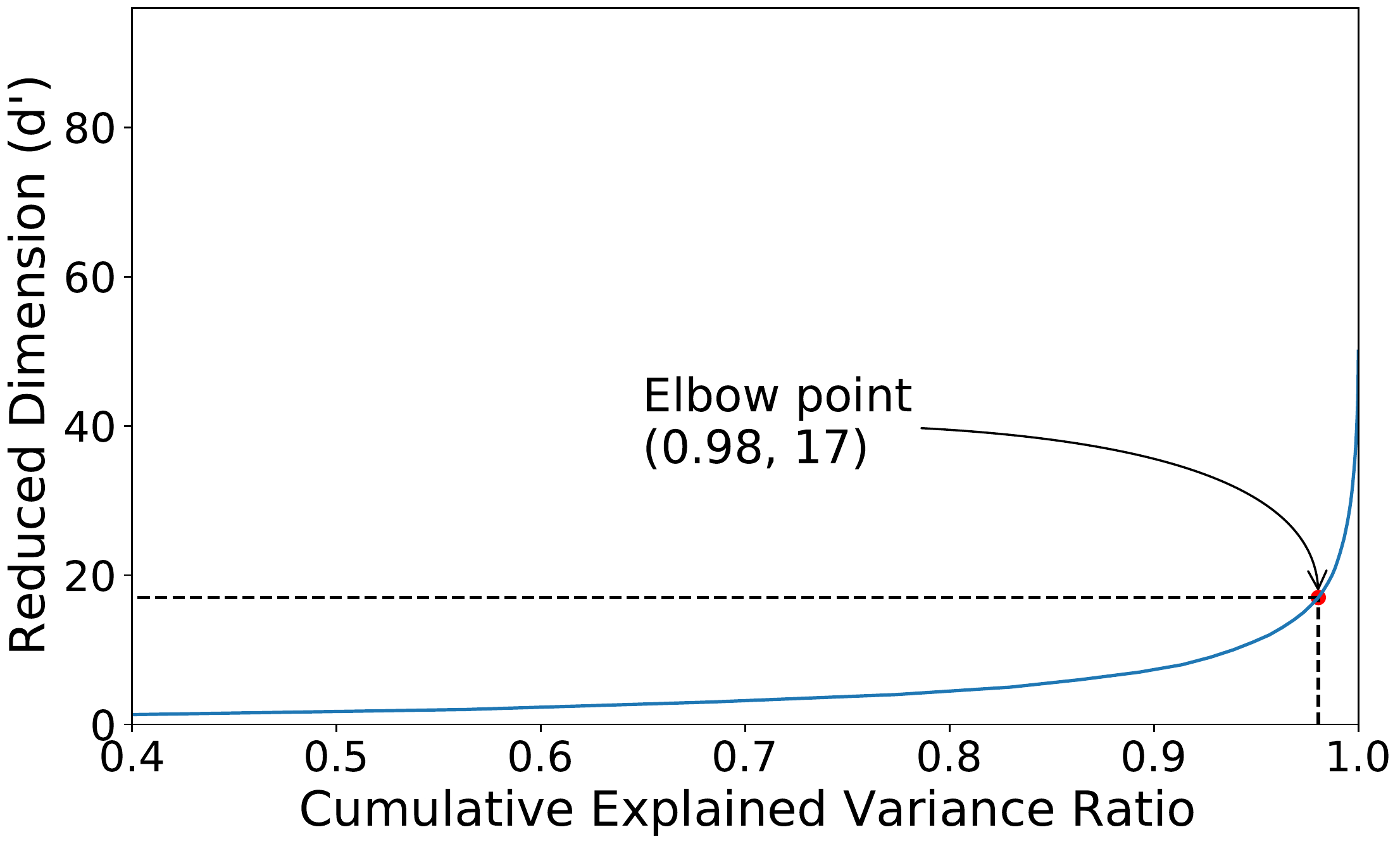}
    \caption{CEVR vs. $d'$ for PCA hyper-parameter setting}
    \label{fig:elbowPCA}
    \vspace{-0.2cm}
\end{figure}

Let the original of the load demand profiles of the households be represented in $\R^d$ space. Thus, the PCA attempts to reduce the dimensionality from $d$-dimensions to $d'$-dimensions while keeping the maximum information intact. This information is represented by the variance of the data. In this regard, PCA computes the eigenvalues and eigenvectors of the data. Here, eigenvalues represent the explained variance across the corresponding eigenvector dimension. After sorting the eigenvectors by the decreasing order of corresponding eigenvalues, PCA takes the top-$d'$ eigenvectors which explain the maximum cumulative variance of the data. This paper uses elbow heuristics~\cite{jain2020clustering} to identify the optimal number of reduced dimensions (say, $d'$). As shown in Fig.~\ref{fig:elbowPCA}, the rate of increment in the cumulative explained variance ratio significantly decreases after reaching the elbow point, and hence the optimal value of $d'$ is chosen to be $17$.

\begin{figure}[!ht]
    \centering
    \includegraphics[width=0.95\columnwidth]{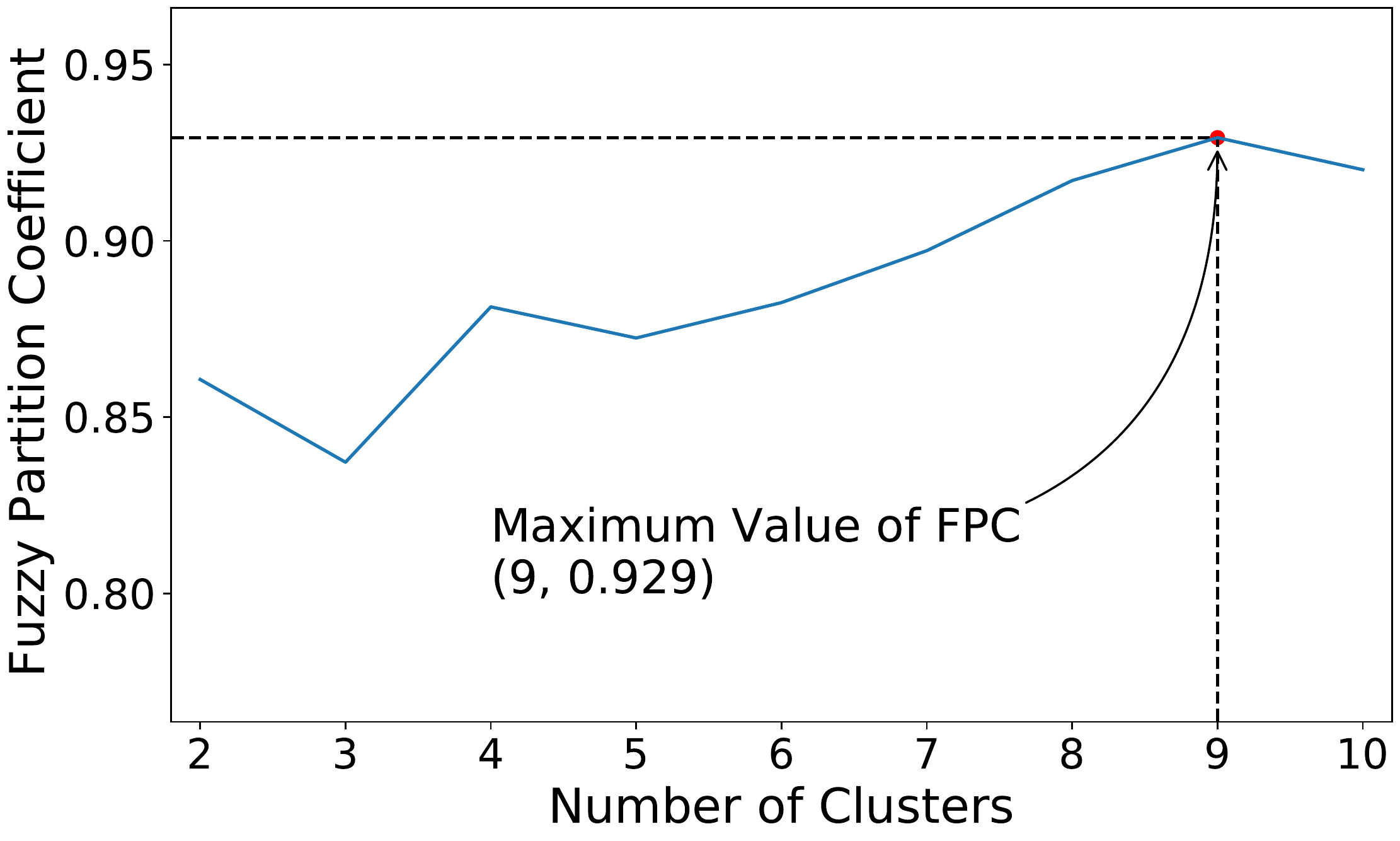}
    \caption{FPC vs. $k$ for FCM hyperparameter setting.}
    \label{fig:fpcFCM}
    \vspace{-0.3cm}
\end{figure}

Fuzzy c-means clustering algorithm needs two hyperparameters to perform clustering. First is the value of fuzzifier which is computed by the method given by Demb\'{e}l\'{e} \& Kastner~\cite{dembele2003}. Second is the number of clusters ($k$) to be generated whose optimal value is computed by computing the value of Dunn's fuzzy partition coefficient (FPC)~\cite{dunn1973} for different number of clusters. The number for which the value of FPC is maximum is chosen as the optimal value (\textit{i.e.} $k=9$) for the second hyperparameter setting (\textit{cf.} Fig.~\ref{fig:fpcFCM}). This study limits the maximum number of clusters to $10$ in accordance with the suggestions made by Al-Otabi~\textit{et~al.}~\cite{al2016} while considering practicality and industrial relevance.

\begin{figure}[!ht]
    \centering
    \includegraphics[trim=40 40 0 0, clip, width=0.95\columnwidth]{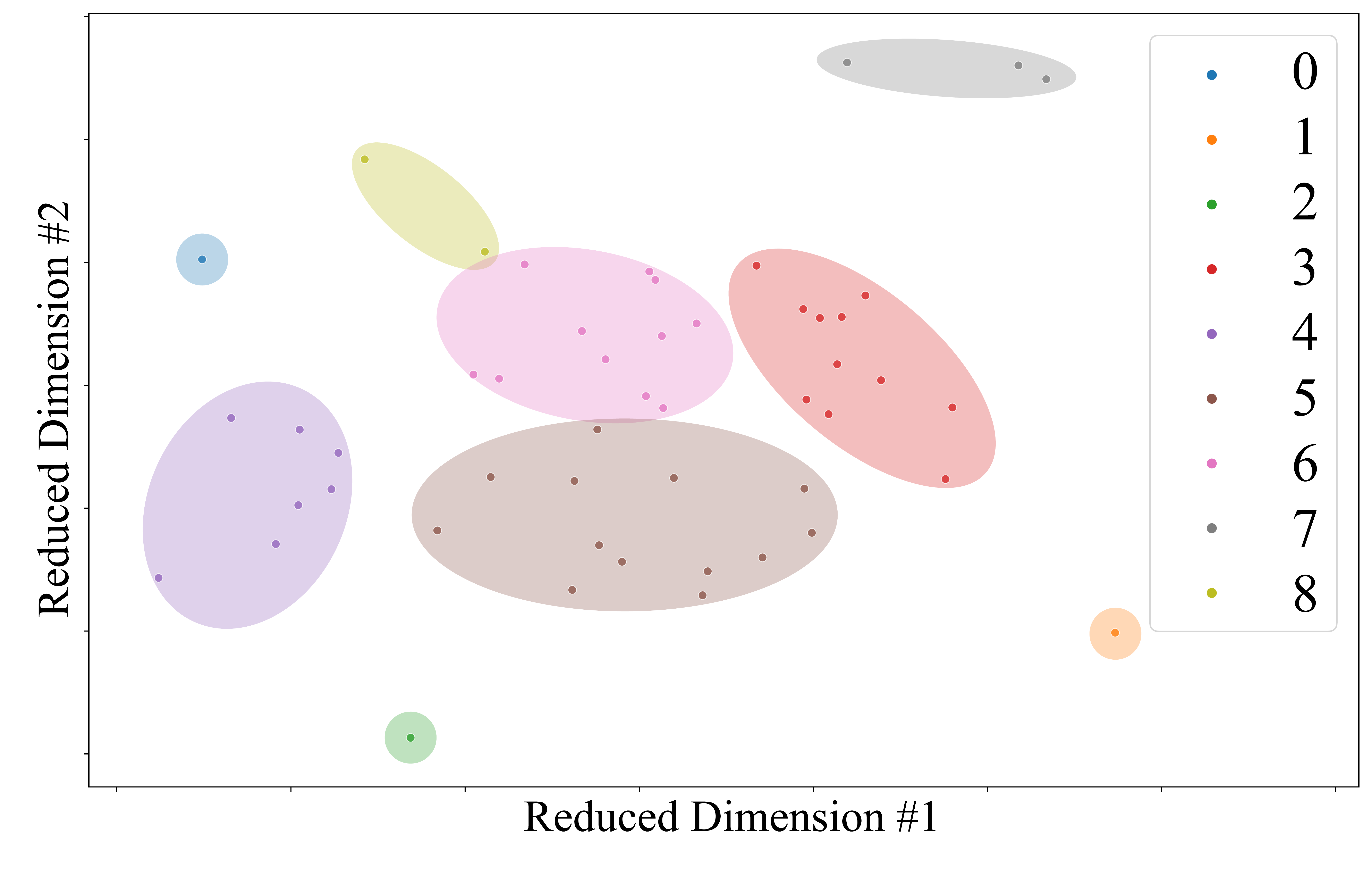}
    \caption{t-SNE plot showing the clustering result using PCA+FCM clustering technique}
    \label{fig:tsne_base}
    \vspace{-0.6cm}
\end{figure}

To check the correctness of clustering results, the t-Distributed Stochastic Neighbor Embedding (t-SNE) is performed to reduce the dimensionality further to $2$. The resulting scatter plot of households (\textit{cf.} Fig.~\ref{fig:tsne_base}) clearly show that the clusters are appropriately dense and well separated. A peculiar aspect of the clusters obtained is that there are three clusters (indexed as $0$, $1$, and $2$ in Fig.~\ref{fig:tsne_base}) that contain a single household.

\subsection{Cluster Validation Indices}\label{sec:CVIs}
This study analyzes only $5$ CVIs that have been most widely used in the domain of clustering residential load demand profiles. These are SH, CH, DB, DI and XB scores as explained hereafter.

\subsubsection{Silhouette Score (SH score)} \label{subsec:SH}
The score uses mean intra-cluster distance ($p$) and mean nearest cluster distance ($q$) to create a measure that give values in the interval $[-1,1]$. A higher SH score reflects better clustering. Scores around $-1$ indicate wrong clustering, values near $0$ depict overlapping clusters and closer to $1$ corresponds to dense and well separated clusters. The major drawback of SH score is that it is biased for bigger clusters~\cite{rousseeuw1987}.

\subsubsection{Calinski-Harabasz Index (CH score)}\label{subsec:CH}
The score is calculated by taking the ratio of sum of the squared distance between cluster center and data center and the summation of the squared distance between all points in cluster and the cluster center. The ratio is then adjusted for the degree of freedom to get the CH score. As the score takes squares into account, it is very much biased towards convex and dense clusters. A higher value of CH score reflects better clustering~\cite{calinski1974}.

\subsubsection{Davies-Bouldin score (DB score)}
\label{subsec:DB}
The DB score measures the average similarity between clusters. The similarity among different clusters is captured by comparing the distance between clusters and their size~\cite{davies1979}.
\begin{equation*}
    DB = \frac{1}{k} \sum_{i=1}^{k} \sum_{j=1}^{k} max \frac{diameter_i+diameter_j}{\text{distance between cluster centroids}} 
\end{equation*}
where $k$ is the number of clusters. Since separation is accounted in the denominator and diameter is in the numerator, a lower DB score indicates better clustering.

\subsubsection{Dunn's Index (DI score)}
\label{subsec:DI}
The DI score is measured by taking the ratio of minimum separation and maximum diameter. It is criticized as a worst-case indicator for not considering the better clusters into account. A higher DI score depicts better clustering~\cite{dunn1973}.

\subsubsection{Xie-Beni Index (XB score)}
\label{subsec:XB}
The score calculates the ratio between compactness and separation. The compactness is measured by the sum of the intra-cluster variances for all the clusters. To deduce separation, the score takes the squared value of the minimum inter cluster distance. Like the DI score, this score also undermines the better separation between clusters~\cite{xie1991}.

All the above-mentioned indices have their own shortcomings. These shortcomings can be identified with experimentation on different cluster characteristics. Since different CVIs are not comparable with each other based on the value they take, a subjective and objective evaluation about their performance is needed for the comparison. It will be carried out in the next section.

\section{Impact of Dataset Characteristics}
\label{sec:Methodology}
The subjective and objective evaluation of different CVIs is carried out for the following data characteristics:

\subsection{Outliers' Impact}
\label{subsec:Outliers}

A single point cluster is not meaningful in any way. The baseline t-SNE plot in Fig.~\ref{fig:tsne_base} shows three such clusters (indexed as 0, 1 and 2). These are essentially the outlier points. Inclusion of such points/clusters in the dataset should not mean anything for the clustering validation exercise and ideally should be left while calculation. But, their addition as a separate cluster reduces the average diameter of all the clusters and increase the average separation. Since most of the CVIs use the concept of diameter and separation to validate the algorithm, they often consider the introduction of outliers as a good thing and favor their inclusion. Their value in the data is overestimated in such cases.

\begin{table*}[!hb]
\setlength\extrarowheight{5pt}
\small
\centering
\caption{The experimental results are derived by using different combinations of outliers in the sample to get the responses of different CVIs. The index of outliers is taken from the individual household clusters in the t-SNE plot of Fig.~\ref{fig:tsne_base}}
\begin{tabular}{ |c|c|c||c|c|c|c|c| }
\hline
Outlier $0$ & Outlier $1$ & Outlier $2$ & SH ($\uparrow$) & CH ($\uparrow$) & DB ($\downarrow$) & DI ($\uparrow$) & XB ($\downarrow$)\\[4pt]
\hline
\xmark & \xmark & \xmark & \textbf{0.21115992} & 16.04573731  & 1.11241178 & \textbf{0.15711491} & 0.74365632\\
\xmark & \xmark & \cmark & 0.20676075 & 14.70077355 & 1.00859215 & \textbf{0.15711491} & 0.72816348\\
\xmark & \cmark & \xmark & 0.20676075 & \textbf{17.98226021} & 0.99171442 & \textbf{0.15711491} & 0.72816348\\
\xmark & \cmark & \cmark & 0.20254115 & 16.57250424 & 0.91595941 & \textbf{0.15711491} & 0.713303  \\
\cmark & \xmark & \xmark & 0.20578034 & 14.63188325 & 1.01062395 & \textbf{0.15711491} & 0.72816348\\
\cmark & \xmark & \cmark & 0.20158074 & 13.67854263 & 0.93250525 & \textbf{0.15711491} & 0.713303\\
\cmark & \cmark & \xmark & 0.20158074 & 16.53295405 & 0.91773724 & \textbf{0.15711491} & 0.713303\\
\cmark & \cmark & \cmark & 0.19754912 & 15.47741121 & \textbf{0.85861914} & \textbf{0.15711491} & \textbf{0.69903694}\\[4pt]
\hline
\end{tabular}

\label{table:outlier_results}
\end{table*}

Since electricity load demand profiles are not yet available at a large scale, such datasets are often prone to having a large number of outliers. Therefore, the study of CVIs in their presence becomes very important. The null hypothesis for this experiment is as follows:

\begin{remarkA}
Addition/Removal of outliers has no effect on CVIs
\end{remarkA}

In order to evaluate this hypothesis, all possible combinations of these three single household clusters are taken. Different validation indexes are then calculated for each possible combination and the CVI responses are measured for assessment.

The SH score improves with the removal of outliers. It can be inferred that the score is biased against the outliers. It considers the presence of single household clusters bad and prefers joining the outliers with the nearest possible clusters. As a result, the diameter of the clusters increases, and the score worsens. It results in violation of the null hypothesis $H^a_0$. 

The CH score behaves in a very peculiar way. It improves whenever the outlier indexed as $1$ is added and reduces with the addition of others. The baseline t-SNE plot in Fig.~\ref{fig:tsne_base} show that the outlier $1$ is placed very far from all other clusters. Whereas, the others are placed relatively close to majority of multi household clusters. The CH score evaluates separation as the squared summation of distance of cluster centroids from the data centroid. The presence of outliers closer to the data centroid reduces average separation and hence the CH score. On the other hand, addition of an outlier far away from the data centroid increases the average separation and hence the CH score as well. Hence, the presence of negative and positive biases in mutually exclusive possibilities for CH score leads to rejection of the null hypothesis $H^a_0$. 

The DI score is immune to the inclusion of single household clusters and hence accepts the null hypothesis by depicting the best possible case. As explained in section~\ref{subsec:DI}, the score takes into account only the worst possible clusters. Outlier clusters belong to the better ones and are ignored by the score. In contrast, the XB and DB indices favor the inclusion of outlier clusters. The indices prefer single household clusters to be taken as separate clusters creating a positive bias for such outlier clusters. The null hypothesis is again rejected by these scores as well.

The study recommends use of DI score when large number of outliers are present. But if all other data characteristics favor any other index, this paper suggests correction of outliers before proceeding with the analysis.

\subsection{Density Impact}
\label{subsec:Density}
\begin{figure}[!ht]
    \centering
    \includegraphics[trim=20 0 0 25, clip, width=0.95\columnwidth]{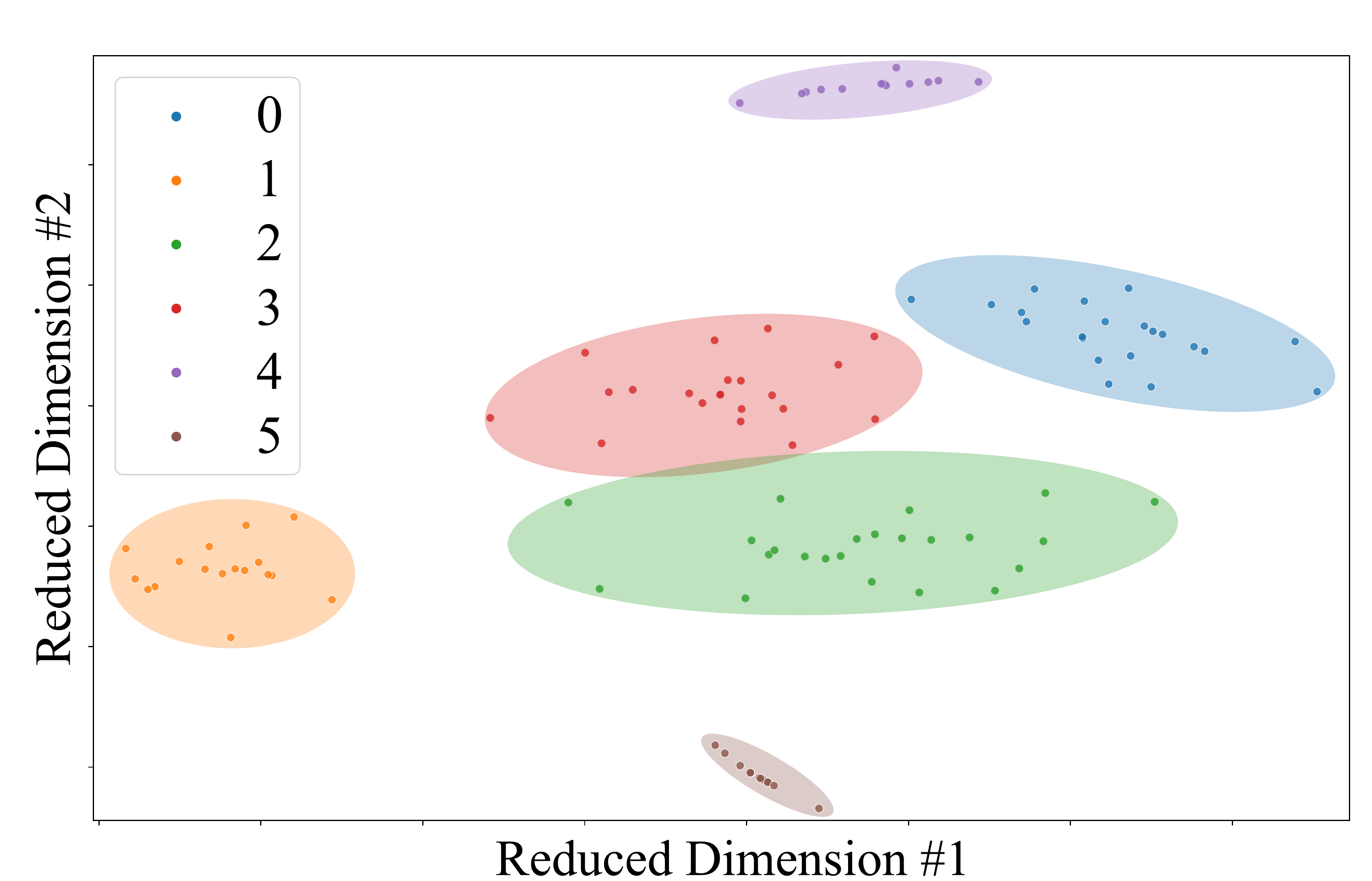}
    \caption{t-SNE plot showing the clustering result for a sample trial using PCA+FCM clustering technique in the increased intra-cluster density experiment.}
    \label{fig:tsne_density}
\end{figure}

Nearly all real-time datasets have observations in clusters with varied densities. This property is especially true with the electricity load demand profiles and hence mandates the analysis. The density experiment is performed by taking the outlier corrected clusters as the baseline. New points are then added in all the clusters to increase the density. They are added in a specific way in new points lie in the \textit{non overlapping cluster region} only. In order to generate such points, a Multivariate Gaussian distribution is fitted for each cluster such that the $std.deviation = \frac{radius}{4}$. As confirmed by the t-SNE plot in Fig.~\ref{fig:tsne_density}, the new points are generated close to the center of the cluster and at the place where maximum observations are rested. Multiple experiments were then performed to get different new points for each experiment. CVIs are then calculated and averaged over to obtain the results. Such exercise will have positive average effect on clustering results without changing the optimal number of clusters. The CVIs are supposed to reflect the positive effects of increased density which will be tested in this experiment.

\begin{remarkB}
Increased intra-cluster density has positive effect on the CVIs
\end{remarkB}

\begin{figure}[!ht]
    \centering
    \includegraphics[width=0.95\columnwidth]{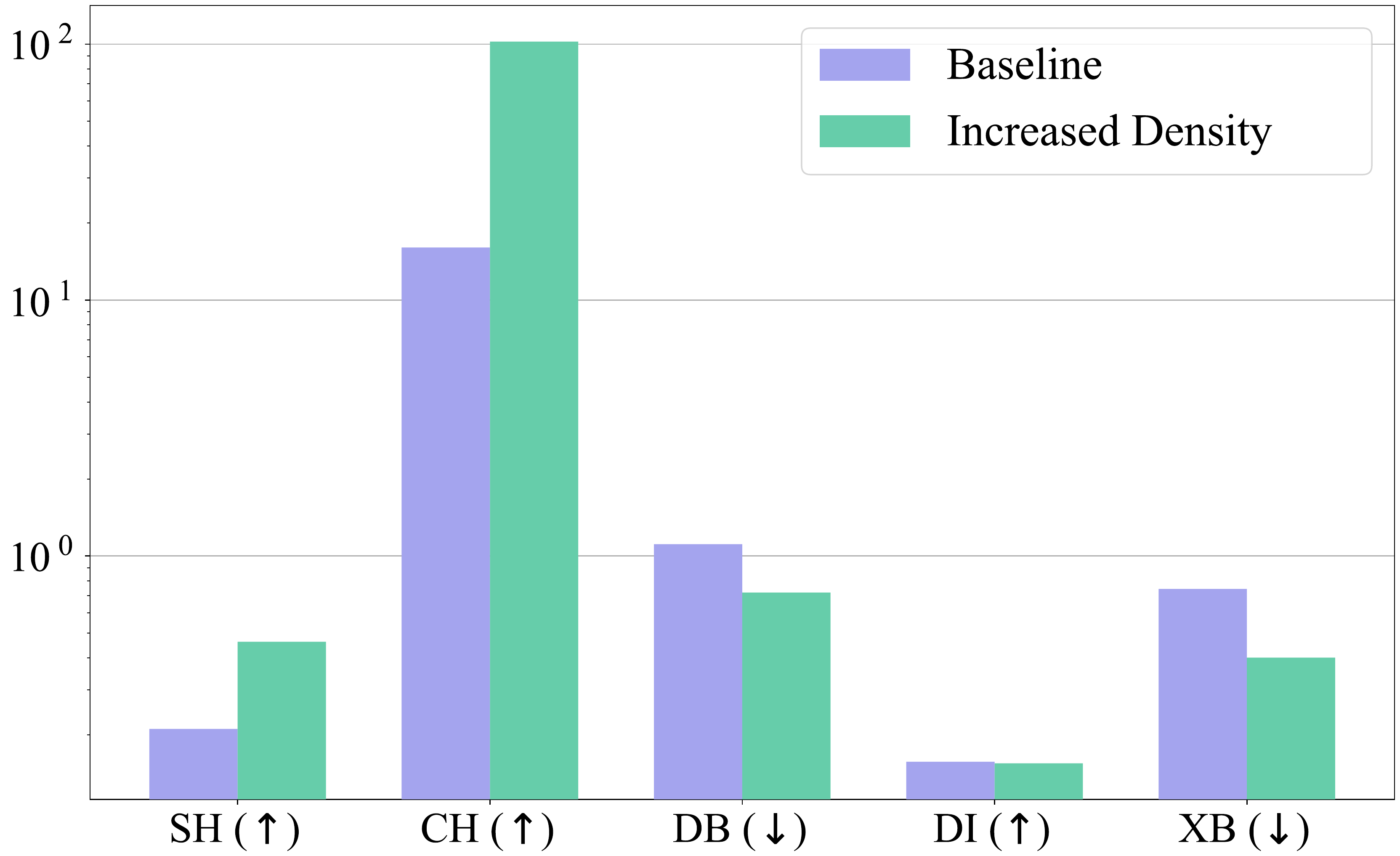}
    \caption{The effect of increased density on value of different cluster validation indices. The CVI scores are averaged over 100 trials because of random addition of points.}
    \label{fig:density_results}
\end{figure}

As depicted in Fig.~\ref{fig:density_results}, increase in density is preferred by all the scores except the DI score. The DI score is positively related to the minimum separation and is negatively related to the maximum diameter. An increase of density in multiple experiments must have marginally influenced the maximum diameter. It is captured by a slight decrease in the average DI score. Since the index doesn't take into account the density of clusters, it recommends an increase in density as bad. On the other hand, as the CH score is very much biased towards convex and dense clusters, overestimation of increased density is observed in our results as well. 

The density results confirm the advantages and limitations of different CVIs. The evidences are against the use of DI score as it rejects the null hypothesis $H^b_0$. Despite acceptance of the null hypothesis by CH score, it's use require caution because of overestimated effect of density. The CH score can easily mask other clustering drawbacks (like outliers, improper clusters etc.) because of the over-positive effect due to highly dense clusters. All the other indices seem to give adequate results.

In conclusion, if there are many households with similar load demand profiles (\textit{i.e.} high expected cluster density), our analysis identifies SH, DB and XB as the most appropriate CVIs. The CH score is over-sensitive and needs to be cautiously used whereas the DI score is not recommended.

\subsection{Effect of Diameter and Separation}

A reduction in the diameter of all the clusters increases the inter cluster separation as well. Since this study use a real electricity load demand profile data, the existing observations cannot be changed. Therefore, an independent study of diameter/separation is not possible. A change in diameter will essentially change separation as well. The experiment studied in this paper proportionately reduces the diameter of all the clusters by twenty percent. The load demand profiles that lie outside the new cluster (but were a part of old cluster) are replaced with the creation of new points in the new cluster. Thus, the density of cluster remains the same. The CVI scores are averaged over $100$ trials because of random addition of points. An increase in separation and reduction in diameter is expected to improve the clustering. Therefore, the null hypothesis is as follows:

\begin{remarkC}
Increased separation and reduced diameter has positive effect on CVIs
\end{remarkC}

\begin{figure}[!ht]
    \centering
    \includegraphics[width=0.95\columnwidth]{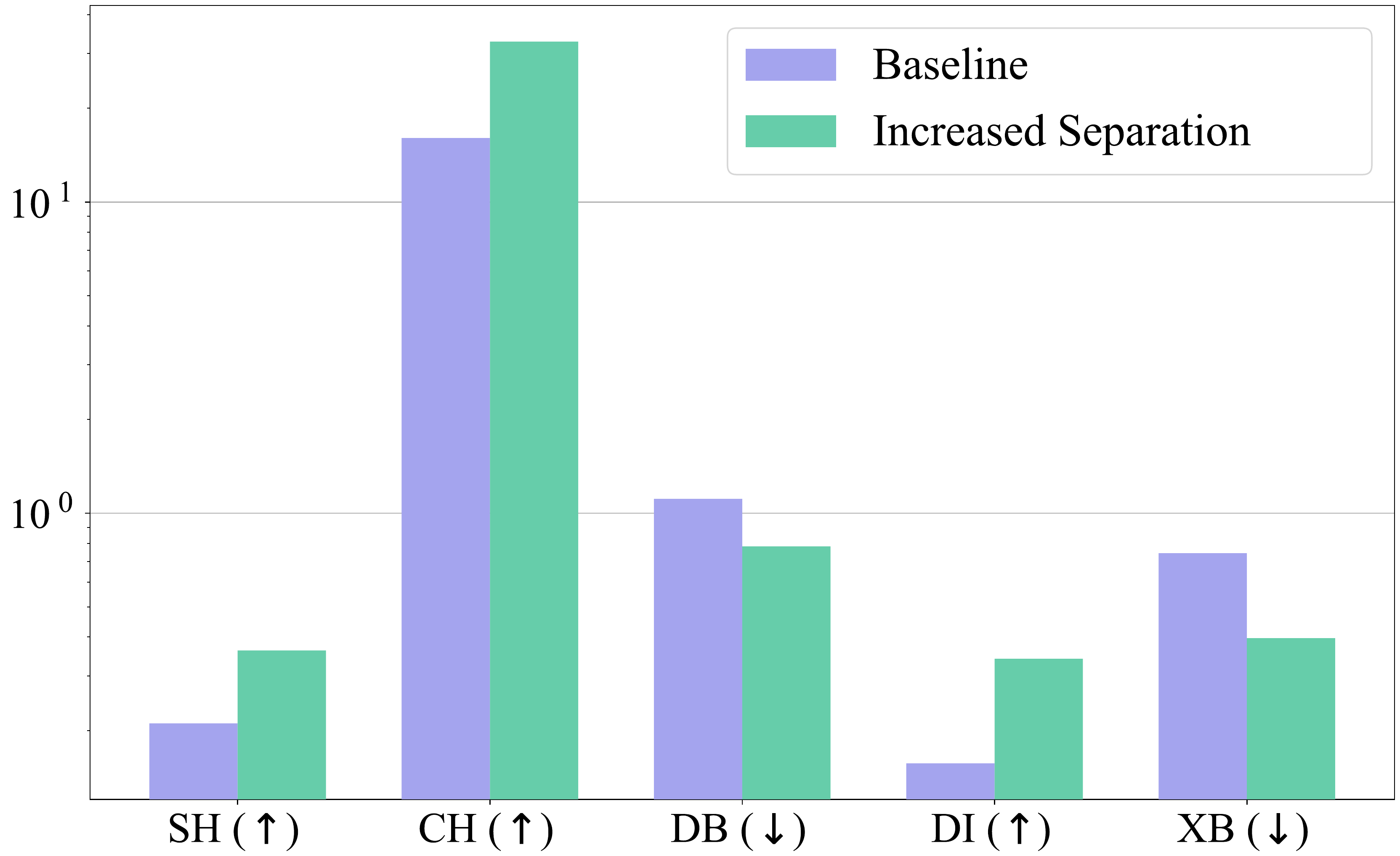}
    \caption{The effect of reduced diameter and increased separation on the value of different cluster validation indices. The CVI scores are averaged over 100 trials because of random addition of points.}
    \label{fig:inc_diameter}
\end{figure}

Fig.~\ref{fig:inc_diameter} shows the results obtained for this experiment. Post reducing the diameter of existing clusters (in the no outlier scenario), all indices show improvement and hence confirm the null hypothesis ($H^c_0$). This means that any of the SH, CH, DB, DI, and XB indices can be used irrespective of the diameter and relative separation of the underlying true clusters in the data. Moreover, since all the considered indices compute cluster diameter and/or separation of clusters at some point, these results also validates the general methodology that is used in this paper.

\section{Conclusion and Future Work}
The paper identifies relevant CVIs based upon different characteristics that are present in a typical dataset of residential load demand profiles. Five validation indices have been studied in this work, namely, the SH score, the CH index, the DB score, the DI score, and the XB Index. If many outliers are present in the data, DI score is the recommended index as it remains unaffected by the addition or removal of outliers. On the other hand, if there are many households with similar load demand profiles (\textit{i.e.} high expected cluster density), our analysis identifies SH, DB and XB as the most appropriate CVIs whereas DI performs the worst. Lastly, it was noted that all the indices behave as expected against the effect of decreased diameter or increased separation between underlying clusters. Therefore, all and any of them can be used in such a scenario.

Furthermore, it should be noted that contrasting results have been obtained for the effects of outliers and differential density. While DI is most recommended index in the first case, its performance is worst (among the $5$ indices considered in this study) in the latter case. This fact strengthens the placement and need of this study in doing subjective analysis of CVIs. In this way, the analysis aims to assist researchers in selecting appropriate CVI(s) while attempting to cluster residential electricity load demand profiles.

In future, we would like to study more CVIs that have been discussed in the literature to compare different clustering results. Furthermore, the extension of this paper will also identify more prevalent dataset characteristics like intra-cluster variance and monotonicity that might impact the performance of one or more CVIs.


\vspace{-0.3cm}


\end{document}